\documentclass[12pt]{iopart}
\usepackage{graphicx}

\begin{document}

\title{Relativistic Multiple Scattering Theory and 
the Relativistic Impulse Approximation}

\author{Khin Maung Maung}

\address{Department of Physics and Astronomy, University of Southern Mississippi, Hattiesburg, Mississippi, 39402, USA}
\ead{khin.maung@usm.edu}

\author{John W. Norbury}

\address{Department of Physics and Astronomy, University of Southern Mississippi, Hattiesburg, Mississippi, 39402, USA and \\ Department of Physics,
Worcester Polytechnic Institute, Worcester, MA 01609}
\ead{john.norbury@usm.edu}

\author{Trina Coleman}

\address{Department of Physics, Hampton University,
   Hampton,VA 23668}
\ead{tlcoleman@howard.edu}

\date{\today}

\begin{abstract}
It is shown  that a relativistic multiple 
scattering theory for hadron-nucleus scattering can be consistently 
formulated in four-dimensions in the context of meson exchange. We  
give a multiple scattering series for the optical potential and discuss
the differences between the relativistic and 
non-relativistic versions.  {\bf } We develop the relativistic multiple scattering series by
separating out the one boson exchange term from the rest of the Feynman series. However this particular separation is not absolutely necessary and we discuss how to include other terms.
We then show   how to
make a three-dimensional reduction for 
hadron-nucleus scattering calculations  and we find that the relative energy 
prescription used in the elastic scattering equation
should be consistent with the one used in the free two-body
$t$-matrix involved in the optical potential.
We also discuss what assumptions are  involved in 
making a Dirac Relativistic Impulse Approximation (RIA).

\end{abstract}


\section{Introduction}

The Relativistic Impulse Approximation (RIA) is one of the most successful tools
in describing hadron-nucleus scattering observables. At the beginning it was
prompted by the success of Dirac phenomenology \cite{clark-first} 
in describing 
proton-nucleus scattering where a parameterized optical potential was used
in a Dirac equation. Soon after this, guided by the non-relativistic $t\rho$
approximation of the optical potential, a relativistic generalization (RIA)
was made \cite{wallace-mcneil}. Over the years
other authors \cite{sink, maung1} 
have also successfully used the RIA with various prescriptions for the
$t$-matrix and the target density. In the case of meson-nucleus scattering, 
the RIA optical potential was successfully used in the Kemmer-Duffin-Patiau 
equation \cite{clark-kdp}. Recently, a Dirac-RIA was used in analyzing
neutron densities \cite{neutron-densities} and nuclear densities 
arising from chiral models \cite{dick}.
which shows that the RIA is also useful in the studies of the
bulk properties of nuclear matter.

The RIA is a very useful tool in medium energy nuclear physics. 
It is based upon the existence of a multiple scattering
theory which obviously  must have some resemblance to non-relativistic
multiple scattering theory. In the non-relativistic theory, there is no
ambiguity in what equation is to be used as the scattering equation. There is
only one equation available, namely the Schrodinger equation. For the NN amplitudes,
there are several possible choices. Some are pure phenomenological fits
and some are calculated from potential models.

In the relativistic case, even the use of Dirac equation in nucleon-nucleus
scattering is questionable. At best, the use of the Dirac equation can be a good
approximation. When the Dirac equation is used in describing the passage of
the projectile nucleon through the nucleus, the tacit assumption is that the
nucleus is infinitely heavy, but in reality it is not. There are also 
ambiguities in choosing the NN amplitude to be used in the RIA 
optical potential, since there are in principle infinitely many relativistic
two-body quasi-potential equations that can be used in producing NN amplitudes.
In order to address these issues, it is important to develop a relativistic
multiple scattering theory (RMST). As 
far as we are aware, there has been only
one attempt to develop an RMST which was done by Maung and Gross \cite{maung-gross, maung}.
In their approach they start from the sum of all meson exchange
diagrams between the projectile and  target nucleus. By considering 
the cancellation between the box and crossed-box diagrams, they concluded that
the projectile-target propagator should be a three-dimensional propagator
with the target on mass-shell when the target is in the ground state. In order
to avoid spurious singularities, Maung and Gross chose the propagator with the projectile
nucleon on mass-shell when the target is in the excited state. They 
developed an RMST and argued that the NN amplitude that should be used
in the RIA optical potential should 
be calculated from a covariant 3-dimensional equation 
with one particle on-mass-shell.

We revisit the formulation of an RMST using a meson exchange model.
Since the cancellation of the box and crossed-box diagrams 
does not work satisfactorily when spin and isospin are 
included, we develop an RMST which is independent of this cancellation.
The paper is organized as follows. We briefly review 
the non-relativistic multiple scattering formalism of Watson \cite{watson}. 
We then develop an RMST for the optical potential from a 
meson exchange model in four-dimensions. 
Also  we discuss what is involved in making the Relativistic Impulse 
Approximation. Finally we discuss the validity of using the Dirac equation
 for proton-nucleus scattering and examine 
the alternatives. 
This paper makes reference only to pion exchange, but in principle any number of different boson exchanges, such as $\sigma$,  $\rho$, $\omega$ etc. could be included. One only has to replace the pion exchange with these other bosons.
 In this paper we emphasize  the multiple scattering formalism and not  the calculation of nucleon-nucleon amplitudes, and hence we do not make any specification of meson-nucleon couplings or  form factors to be used.  In the literature numerous authors over the years have used different relativistic equations and meson-nucleon couplings and various types of form factors have been employed in nucleon-nucleon phenomenology.

\section{Review of non-relativistic theory}
\begin{figure}[!h]
\begin{center}
 \includegraphics[width=6cm]{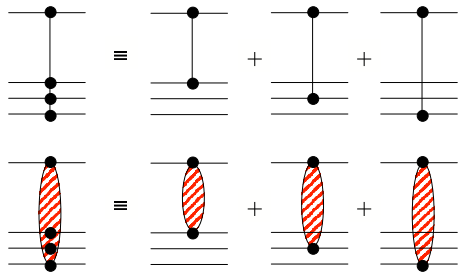}
 \end{center}
\caption{Diagrams  of  the definitions
$V \equiv  \sum\limits_{i=1}^A v_{0i}$  (top) and  $T \equiv  \sum\limits_{i=1}^A T_{0i}$ (bottom) for a target containing 3 particles.}
\label{figone}
\end{figure}

This section contains a  review of non-relativistic theory following references \cite{maung1, maung,  tandy, eisenberg, scheck, maung2}, which provide an  introduction to the topics of non-relativistic  \cite{tandy, eisenberg, scheck}  and relativistic   \cite{maung1, maung, tandy,  maung2} multiple scattering theory.  
This review is included so that the reader can more easily understand  the new relativistic multiple scattering theory introduced later in the paper.
The full  $pA$ hamiltonian is given by
\begin{eqnarray}
H= H_0+V = h_0 + H_A +V
\end{eqnarray}
where $h_0$ is the kinetic energy operator of the projectile and $H_A$ is the full $A$-body hamiltonian of the target.  $H_A$ contains all the target nuclear structure information with
\begin{eqnarray}
 H_A = \sum\limits_{i=1}^{A} h_i +  \sum\limits_{i < j}^A v_{ij}
\end{eqnarray}
This target Hamiltonian is just the  sum of the target nucleon kinetic energies plus the sum of their pair interactions \cite{tandy}.
The residual interaction $V$ is given by the sum of the interactions between the projectile and target particles,
\begin{eqnarray}
V=  \sum\limits_{i=1}^A v_{0i} \label{sum}
\end{eqnarray}
where $v_{0i}$ denotes the interaction between the projectile, labeled particle ``0" and the target nucleon labeled with index ``$i$".
We also write the   $T$-matrix,
\begin{eqnarray}
T &\equiv &  \sum\limits_{i=1}^A T_{0i}
\end{eqnarray}
This and equation (\ref{sum}) are  shown using  diagrams defined in Figure   \ref{figone}  and defined in this way, the diagrams themselves can then be easily iterated, as shown  later.

\begin{figure}[!h]
\begin{center}
 \includegraphics[width=6cm]{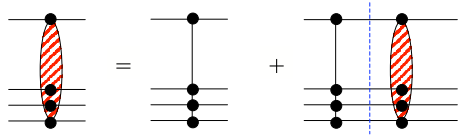}
\end{center}
\caption{Diagram  of  the Lippman-Schwinger equation,
$T  = V +VG_0T$, for nucleon-nucleus scattering. The propagator $G_0=  \frac{1}{E-h_0 - H_A +i\eta} $ is shown by the vertical dashed line, which goes through all of the nucleons in the target because the propagator contains the full nuclear hamiltonian $H_A$.}
\label{figtwo}
\end{figure}

\noindent The eigenstates (i.e. nuclear bound states) of the nuclear target Hamiltonian $H_A$ satisfy
\begin{eqnarray}
H_A |\phi_n^A  \rangle =E_n  |\phi_n^A \rangle
\end{eqnarray}
From the
 beginning the A-body problem is separated from the rest and we assume that
there is  some means of obtaining the solution of this A-body bound state
problem.
The projectile scattering eigenstates satisfy 
\begin{eqnarray}
h_0 |{\bf k}  \rangle =E_k  | {\bf k} \rangle
\end{eqnarray}
The eigenstates of the full unperturbed $pA$ Hamiltonian $H_0=h_0+H_A$ satisfy
\begin{eqnarray}
H_0 |\Phi \rangle =E |\Phi \rangle  \label{kepluseigen}
\end{eqnarray}
where the energy $E$ is total kinetic energie of the projectile and  target plus the eigenenergies of the target. In the lab frame the target kinetic energy is zero.
The  initial and final states  are
\begin{eqnarray}
 |\Phi_i  \rangle  =   |\phi_i  \rangle   | {\bf k}_i \rangle \;,
 \hspace*{1.5cm}
  |\Phi_f  \rangle  =   |\phi_f  \rangle   | {\bf k}_f \rangle
\end{eqnarray}
The transition amplitude between different intial and final states of the same energy is
\begin{eqnarray}
T_{fi} \equiv  \langle \Phi_f |T| \Phi_i \rangle   \label{amp}
\end{eqnarray}
with the $T$ matrix operator given by the   Lippman-Schwinger equation (LSE),
\begin{eqnarray}
T=V+VG_0T  \label{lse}
\end{eqnarray}
where the free propagator of the $pA$ system is
\begin{eqnarray}
G_0 = \frac{1}{E-H_0 +i\eta} =  \frac{1}{E-h_0 - H_A +i\eta} 
\end{eqnarray}
 The diagrammatic representation of  the LSE
is shown in Figure  \ref{figtwo}. Note that there are three energies involved in the evaluation of $\langle {\bf k}_f |T(E)  |  {\bf k}_i \rangle$, where the energy $E$ in $T(E)$ is the energy appearing in $G_0$ above and in equation (\ref{kepluseigen}).
There is also  the initial energy of the projectile $E_i$ and the final energy $E_f$ of the scattered projectile  and any emitted particles. If the three energies are all different the process is described as 
 completely off-energy-shell \cite{eisenberg} and we have the  completely off-energy-shell
 $T$-matrix  $\langle {\bf k}_f |T(E)  |  {\bf k}_i \rangle$ \cite{eisenberg}. We can also define two half
 off-energy-shell $T$-matrices as  $\langle {\bf k}_f |T(E_i)  |  {\bf k}_i \rangle$ or $\langle {\bf k}_f |T(E_f)  |  {\bf k}_i \rangle$ when $E=E_i$ or $E=E_f$. These three amplitudes become equal in the completely on-energy-shell situation where $E=E_i=E_f$ \cite{eisenberg}.

\subsection{First order multiple scattering}

\begin{figure}[!h]
\begin{center}
\includegraphics[width=9cm]{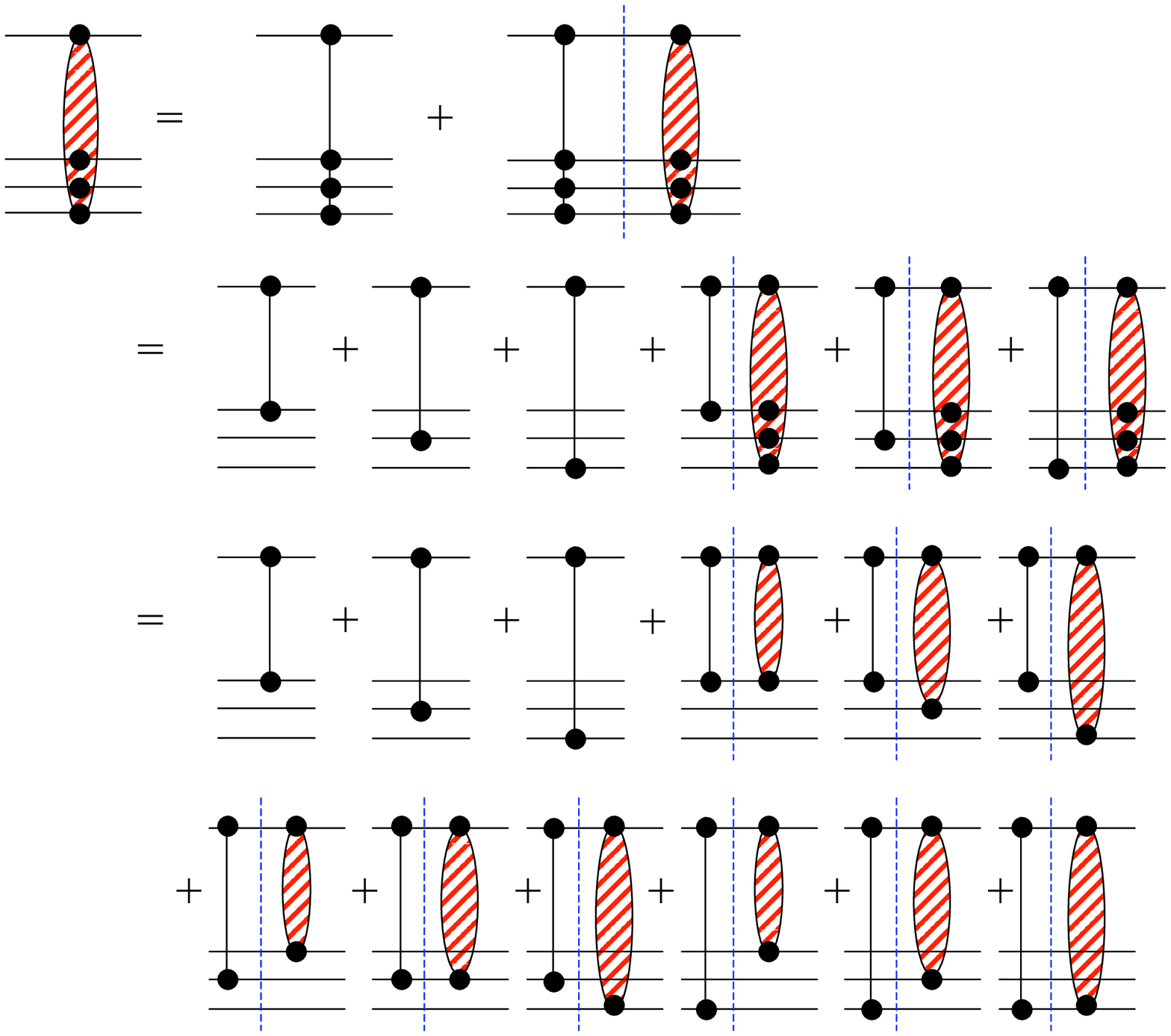} 
 \end{center}
\caption{Diagram  of  the Lippman-Schwinger equation and its expansion,\\
$T  = V +VG_0T
 = \sum\limits_{i=1}^Av_{0i} + \sum\limits_{i=1}^A v_{0i} G_0  T
 = \sum\limits_{i=1}^Av_{0i} + \sum\limits_{i=1}^A v_{0i} G_0  \sum\limits_j T_{0j} $
 }
\label{figthree}
\end{figure}

\begin{figure}[!h]
\begin{center}
\includegraphics[width=7cm]{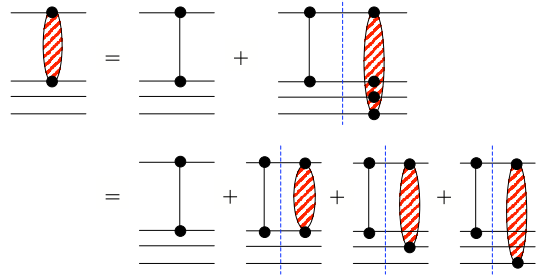}
 \end{center}
\caption{Diagram  of  the $i^{\rm th}$ term of the Lippman-Schwinger equation and its expansion,
$T_{0i}  = v_{0i}+v_{0i}G_0T =v_{0i} + v_{0i} G_0  \sum\limits_j T_{0j}$ }
\label{figfour}
\end{figure}

 Substituting (\ref{sum}) into (\ref{lse}) gives
what we call the  Lippman-Schwinger expansion,
\begin{eqnarray}
T=  \sum\limits_{i=1}^A v_{0i} +  \sum\limits_{i=1}^A v_{0i} G_0T   \label{lseexpansion}
\end{eqnarray}
with the $i^{\rm th}$ term
\begin{eqnarray}
T_{0i}  &\equiv & v_{0i}  + v_{0i}  G_0 T   =  v_{0i}  + v_{0i}  G_0   \sum\limits_j T_{0j}  \label{this}
\end{eqnarray}
which, upon iteration gives \cite{tandy}
\begin{eqnarray}
T_{0i} \equiv v_{0i}  + v_{0i}  G_0   \sum\limits_j v_{0j} + \cdots  \label{one}
\end{eqnarray}
Suppose the target is a nucleus  with three nucleons. Then this expression is
\begin{eqnarray}
T_{0i}  &\equiv&  v_{0i}  + v_{0i}  G_0  ( v_{01} +v_{02}+v_{03}) + \cdots
\end{eqnarray}
where the first term represents a single interaction between the projectile and the $i$-th
target nucleon. The collection of second terms represent a double interaction between the projectile and the $i$-th target nucleon. This consists of a single  interaction between the projectile and the $i$-th target nucleon, followed by propagation represented by $G_0$ and then another single interactions between the projectile and {\em each} of the target nucleons. Figures  \ref{figthree} and \ref{figfour}  show the series for  a proton scattering from a  nucleus  with three nucleons.   One can see that the diagram definitions in Figure  \ref{figone}  allows for the diagrams themselves to be iterated as in Figures  \ref{figthree} and \ref{figfour}.

\begin{figure}
\begin{center}
\includegraphics[width=6cm]{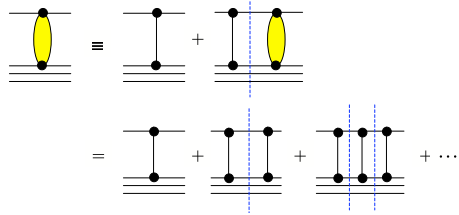}
 \end{center}
\caption{Diagram  of the single scattering term,
$
t_{0i} \equiv v_{0i}  + v_{0i}  G_0   t_{0i}   
= v_{0i}  + v_{0i}  G_0   v_{0i} + v_{0i}  G_0    v_{0i}  G_0   v_{0i} + \cdots
$ }
\label{figfive}
\end{figure}

 \noindent
Each higher order  term in (\ref{one}) contains terms where the interaction occurs multiple times on the same target nucleon. These can be separated off by writing
\begin{eqnarray}
T_{0i} &=& v_{0i}  + v_{0i}  G_0   v_{0i} + v_{0i}  G_0    v_{0i}  G_0   v_{0i} \cdots  
+ v_{0i}  G_0   \sum\limits_{j \neq i} v_{0j} + \cdots   \\
&= & t_{0i} +   v_{0i}  G_0   \sum\limits_{j \neq i} v_{0j} + \cdots  
\end{eqnarray}
with (see Figure  \ref{figfive})
\begin{eqnarray}
t_{0i} &\equiv &  v_{0i}  + v_{0i}  G_0   t_{0i} 
= v_{0i}  + v_{0i}  G_0   v_{0i} + v_{0i}  G_0    v_{0i}  G_0   v_{0i} + \cdots    \label{series}
\end{eqnarray}

\noindent Write equation (\ref{this}) as
\begin{eqnarray}
T_{0i}  &\equiv & v_{0i}  + v_{0i}  G_0 T  = v_{0i}  + v_{0i}  G_0   \sum\limits_j T_{0j}  \\
&=& v_{0i}  +  v_{0i}  G_0   T_{0i}  + v_{0i}  G_0   \sum\limits_{j \neq i}  T_{0j} 
\end{eqnarray}

 \noindent
Rearrange as 
\begin{eqnarray}
(1- v_{0i}  G_0) T_{0i}  &= & v_{0i}   + v_{0i}  G_0   \sum\limits_{j \neq i}  T_{0j}  
\end{eqnarray}
giving
\begin{eqnarray}
T_{0i}&=& (1- v_{0i}  G_0)^{-1} v_{0i}   +  (1- v_{0i}  G_0)^{-1} v_{0i}  G_0   \sum\limits_{j \neq i}  T_{0j} 
\end{eqnarray}
Using the binomial series 
$\frac{1}{1-x} = 1+x +x^2  +\cdots $
gives
\begin{eqnarray}
 (1- v_{0i}  G_0)^{-1} v_{0i}   &=& v_{0i} + v_{0i}G_0v_{0i} +v_{0i}G_0v_{0i}G_0v_{0i} +\cdots 
 =  t_{0i}
\end{eqnarray}
to finally give the  Watson multiple scattering series  \cite{watson, tandy, eisenberg}
\begin{eqnarray}
 T_{0i}  &= & t_{0i}   + t_{0i}  G_0   \sum\limits_{j \neq i}  T_{0j}    \label{watsoneqn}
\end{eqnarray}
The advantage of this series is that it is an expression for the full $T$ matrix involving scattering amplitudes $t_{oi}$ rather than potentials $v_{oi}$, with each $t_{oi}$  containing an  infinite number of the $v_{oi}$ terms.

\subsection{Single scattering approximation (SSA)}
\begin{figure}[!h]
\begin{center}
\includegraphics[width=9cm]{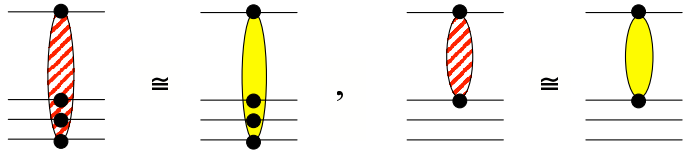}
 \end{center}
\caption{Diagram of the single scattering approximation
$T \approx  \sum\limits_{i=1}^A t_{0i}$ (left diagram) or  the $i^{\rm th}$ term  $T_{0i} \approx  t_{0i}$ (right diagram). }
\label{figsix}
\end{figure}

The  single scattering approximation is
\begin{eqnarray}
T_{0i} \approx t_{0i} 
\end{eqnarray}
so that (\ref{lseexpansion}) becomes
\begin{eqnarray}
T  \equiv  \sum\limits_{i=1}^A T_{0i} \approx   \sum\limits_{i=1}^A t_{0i}  = t_{01} + t_{02} + t_{03} + \cdots  \label{ssa}
\end{eqnarray}
The  single scattering approximation is shown in Figure  \ref{figsix}, and  ``may be valid for weak scattering  or for dilute systems. This works for electron scattering" \cite{eisenberg}.  Tandy \cite{tandy} mentions that the  SSA ``makes a great deal of sense, since the projectile, once it comes close to a given target particle may multiply interact with that particle, but once it is ejected will, with a high degree of probability, ``miss" all the other target particles."

\subsection{Impulse approximation (IA)}

Tandy explains the SSA as follows \cite{tandy}. ``The required amplitude described by $t_{0i}$ does not correspond to the solution of a (free) nucleon-nucleon scattering problem. Because of the presence of $H_A$ in the Green's function operator $G_0$ of equation (\ref{series}), the motion of nucleon $i$ is governed not only by its interaction $v_{0i}$ with the projectile, but also by its interaction with the other constituents of the target. A further approximation can be envisaged in which $H_A$ is assumed to simply set an energy scale so that the solution of equation (\ref{series}) might be replaced by the solution of a free nucleon-nucleon scattering problem. With this interpretation of $t_{0i}$, equation (\ref{ssa}) is referred to as the impulse approximation." 
Thus there  are {\em two} pieces to the single scattering IA
 The first piece consists of the SSA but with the replacement
 \cite{eisenberg}
\begin{eqnarray}
t_{0i} \approx t_{0i}^{\rm free}
\end{eqnarray}
and the second piece consists of using the free Green function
\begin{eqnarray}
G_0 \equiv \frac{1}{E-h_0-H_A} \;\;  \approx \; \; G_0^{\rm free} \equiv  \frac{1}{E-h_0-h_i}  \label{gia}
\end{eqnarray}
This essentially means that the target nucleus is treated as though it is not bound.

\subsection{Optical potential and Watson series}

For elastic scattering it is useful to use an optical potential
 which reduces the orginal many-body elastic scattering problem to a one-body
 problem. All the complicated many-body problems are now included in the
 optical potential. Therefore for practical calculations
 approximations have to be made Êto determine the optical potential to be
 used in the scattering equation. 
 We follow Feshbach \cite{tandy, feshbach1, feshbach2} and define a ground state projector P and an
 operator Q which projects onto the complementary
 space of the excited target states including inelastic break-up states   \cite{tandy, feshbach1, feshbach2}  so that
\begin{eqnarray}
P+Q=1
\end{eqnarray}
where the projector of the target ground state is
\begin{eqnarray}
P \equiv |\phi_0\rangle \langle \phi_0 |
\end{eqnarray}
with $|\phi_0\rangle$ denoting the target nuclear ground state, giving
\begin{eqnarray}
P   |\phi_\alpha \rangle  = |\phi_0\rangle \langle \phi_0    |\phi_\alpha \rangle
= |\phi_0\rangle \delta_{\alpha 0} = |\phi_0  \rangle
\end{eqnarray}
 Now for  elastic scattering the initial and final states are the ground state \cite{tandy}, namely
\begin{eqnarray}
|\phi_i\rangle_{\rm elastic} = |\phi_f\rangle_{\rm elastic} = |\phi_0\rangle
\end{eqnarray}
so that 
\begin{eqnarray}
T_{fi \;\rm elastic} \equiv  \langle \Phi_f |T| \Phi_i \rangle_{\rm elastic}  =\langle{\bf k}_f| \langle \phi_f |T| \phi_i \rangle |{\bf k}_i \rangle
&=&\langle{\bf k}_f|  \langle \phi_0 |T| \phi_0 \rangle|{\bf k}_i \rangle   \\
&=& \langle \Phi_f |PTP| \Phi_i \rangle   
\end{eqnarray}
Thus for elastic scattering
\begin{eqnarray}
T_{\rm elastic} \equiv  PTP
\end{eqnarray}
In analogy with the LSE (\ref{lse}), define the optical potential as \cite{tandy}
\begin{eqnarray}
PTP \equiv PUP + PUP G_0 PTP
\end{eqnarray}
or
\begin{eqnarray}
T  &\equiv & U + UP G_0 T \label{optical} \\
U&=&V+VG_0QU \label{micro}
\end{eqnarray}
This will help us obtain  the microscopic content of the optical potential. Equations (\ref{optical}) and 
(\ref{micro}) are completely equivalent to the Lippman-Schwinger equation (\ref{lse}). This is easily seen by writing  $U=(1-VG_0Q)^{-1}V$ and substitute into (\ref{optical}). Multiply the new (\ref{optical}) by   $1-VG_0Q$ and (\ref{lse}) results.
Because $P^2=P$ and $Q^2=Q$  and  $P$, $Q$ both commute with $G_0$, then instead of  (\ref{optical}) and (\ref{micro})  we can define $U$ differently and write
\begin{eqnarray}
T  &\equiv &  U + UP G_0P T  \label{khindefn} \\
U &=& V+VQG_0QU   \label{khindefn1}
\end{eqnarray}
These equations are also completely equivalent to the Lippman-Schwinger equation (\ref{lse}).  We shall use the above two equations, instead of  (\ref{optical}) and (\ref{micro})  from now on.
Following definition (\ref{lseexpansion}) we now define
\begin{eqnarray}
U\equiv  \sum\limits_{i=1}^A U_{0i}
\end{eqnarray}
and similar to equation (\ref{this}), we have
\begin{eqnarray}
U_{0i}  &=& v_{0i}  + v_{0i}   Q  G_0  Q U  
= v_{0i}  + v_{0i}    Q  G_0 Q   \sum\limits_j U_{0j}  
\end{eqnarray}
Now  define  \cite{watson, francis}  an operator $\tau_i$  
\begin{eqnarray}
\tau_{0i} &\equiv &  v_{0i}  + v_{0i}    Q  G_0Q   \tau_{0i}   \label{tau}
\end{eqnarray}
which is analogous to (\ref{series}). Therefore
we get the   Watson multiple scattering series for the optical potential \cite{watson, tandy, francis}
\begin{eqnarray}
U_{0i}  
&=&  \tau_{0i}  + \tau_{0i}   Q   G_0 Q   \sum\limits_{j \neq i}^A U_{0j}  
\end{eqnarray}
analogous to (\ref{watsoneqn}). Summing gives 
\begin{eqnarray}
U &=&  \sum\limits_{i=1}^A \tau_{0i} +  \sum\limits_{i=1}^A \tau_{0i}    Q  G_0 Q \sum\limits_{j\neq i}^A  U_{0j}    
= \sum\limits_{i=1}^A \tau_{0i} +  \sum\limits_{i =1}^A \tau_{0i}   Q   G_0 Q \sum\limits_{j\neq i}^A\tau_{0j} + \cdots  \label{watsonoptical}  
 \end{eqnarray}
One may ask why we went to all this trouble to develop an optical theory.  Why don't we just calculate the ground state $T$-matrix element  $\langle 0|T|0 \rangle$?  We could calculate matrix elements using either the Lippman-Schwinger expansion in equation (\ref{lseexpansion}) or the Watson series in equation (\ref{watsoneqn}). The trouble is that both equations involve $G_0$, which  we have seen involves a sum over all excited states, which makes the LSE very difficult to solve.
However with $T$ expressed in terms of $U$ in equation (\ref{khindefn}), we see that it contains the term $PG_0P$ which means that it only includes intermediate states with the target in the
 ground state.
The single scattering 
approximation or the first order optical potential is obtained by keeping
the first term only. The successive terms
can be interpreted as the double scattering term, triple scattering terms etc.
and hence the name multiple scattering.
The first order Watson optical potential is \cite{maung}
\begin{eqnarray}
U &=&  \sum\limits_{i=1}^A \tau_{0i} 
\end{eqnarray}
but $\tau$ is {\em not} the free two-body $t^{\rm free}$ matrix because of the presence of the many body propagator $QG_0Q$ in (\ref{tau}), where all intermediate states are in excited states.  For practical calculations a free two-body $t$-matrix is more easily available.  The free two-body $t$ matrix is defined
 \begin{eqnarray}
t^{\rm free}_{0i} \equiv v_{0i} +v_{0i}gt^{\rm free}_{0i}  \label{free}
\end{eqnarray}
where $g$ is the free two body propagator. The relation between $t^{\rm free}$ and the Watson operator
 $\tau$ is 
 \begin{eqnarray}
\tau_{0i} =  t^{\rm free}_{0i} + t^{\rm free}_{0i}(QG_0Q-g) \tau_{0i}  \label{relation}
\end{eqnarray}
For high projectile energies one usually approximates $\tau$ by $t^{\rm free}$ (impulse approximation) and obtains the  first order Watson impulse approximation optical potential
\begin{eqnarray}
U^{1\rm st}_{\rm impulse} &=&  \sum\limits_{i=1}^A t^{\rm free}_{0i} 
\end{eqnarray}
The  Watson optical potential  in terms of the free two body $t$ matrix is usually written  
\begin{eqnarray}
U &=&  \sum\limits_{i=1}^A t_{0i}^{\rm free} +  \sum\limits_{i=1}^A t_{0i}^{\rm free} (Q G_0 Q -g)U_{0i}+
 \sum\limits_{i=1}^A \sum\limits_{j\neq i}^A  t_{0i}^{\rm free} QG_0Q U_{0j}  \\
 &= & \sum\limits_{i=1}^A t_{0i}^{\rm free} +  \sum\limits_{i=1}^A t_{0i}^{\rm free} (Q G_0 Q -g)t^{\rm free}_{0i}+
 \sum\limits_{i=1}^A  t_{0i}^{\rm free} QG_0Q   \sum\limits_{j\neq i}^A t^{\rm free}_{0j} + \cdots
 \label{correction}
 \end{eqnarray}
up to second order. 
Obviously the first term is the single scattering term, the second term is
the single scattering propagator correction term and 
the third term is the double
scattering term etc. The first term alone gives the 
single scattering or the first-order Impulse Approximation (IA) optical
potential operator. It is important to note that it is 
not the same as approximating the Watson $\tau$ with
free $t$-matrix at the single scattering level. As can be seen from the
above equation there also is a propagator correction term at the single
scattering level although it is second order in $t$. Actually the propagator 
correction term exists to all orders for each level of scattering, i.e.
for single scattering, double scattering etc. The propagator
correction term can be interpreted as the medium correction term since
it corrects the use of free propagator instead of the propagator with
the excited intermediate target state.  
For high projectile energies the differences between $\tau$ and $t^{\rm free}$ become negligible. The last term represents the multiple scattering. For non-relativistic calculations 
 $t^{\rm free}$ can be obtained from (\ref{free}) by using a choice of $v$ such as the Reid potential.

\section{Relativistic multiple scattering}

\begin{figure}[!h]
\begin{center}
\includegraphics[width=11cm]{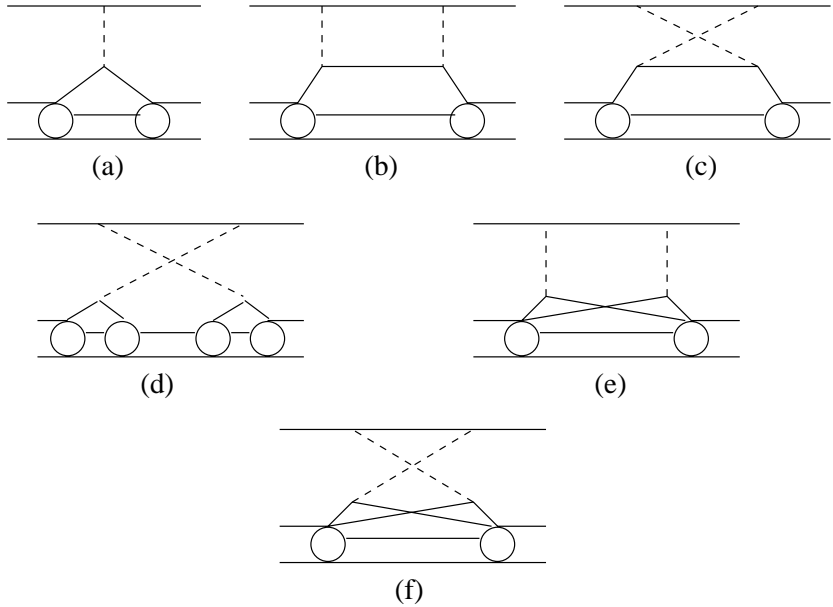}
 \end{center}
\caption{Diagrams in the kernel of the nucleon-nucleus 
Bethe-Salpeter equation
up to the fourth order in the coupling.  The projectile is represented by a single line and the target is represented by a double line.}
\label{figseven}
\end{figure}

Now we discuss a formulation of an RMST in the context of
meson exchange. That is, the interaction between the projectile and the
A-body target nucleus will be mediated by meson exchange.
We start from the fact that the
$t$-matrix for the relativistic projectile-target scattering
is given by the Bethe-Salpeter equation where the kernel is the sum of all
two-body (projectile and the A-body target nucleus) irreducible diagrams. 
The derivation of a multiple scattering series from a field
 theoretical Lagrangian is a very difficult and open problem.
 We want to develop
a multiple scattering theory from the meson exchange point of view and 
want to see what approximations are involved in the RIA. Therefore in all
the diagrams, all self energy and vertex corrections are included as 
renormalized masses and vertices with form factors. The 
kernel of the equation is denoted by $V$ and diagrams up to the fourth
order in the meson-nucleon coupling are shown in Figure  \ref{figseven}.

The Bethe-Salpeter equation for the scattering is 
\begin{equation}
T=V+VG_0T  \label{bse}
\end{equation}
where $G_0$ is the four-dimensional two-body propagator of the 
projectile-target system. The first term in $V$ shown in 
Figure  \ref{figseven}a is the sum of one boson
exchange  interactions between the projectile and the target nucleons.
We label these by $\sum\limits_i K_{0i}$. The second and the third diagrams shown in 
Figure  \ref{figseven}b and \ref{figseven}c are the two meson exchange diagrams between the projectile
and the $i^{\rm th}$ target nucleon and we will denote them by
$\sum\limits_i K_{ii}^{(1)}$ and $\sum\limits_i K_{ii}^{(2)}$.  In a similar manner
we will denote third and higher order
diagrams involving multi-meson exchange between the projectile and a single
target nucleon by $\sum\limits_i K_{iii}^{(m)}$, $\sum\limits_i K_{iiii}^{(m)}$ etc.
The box diagrams are labeled with $m=1$ and the cross box diagrams are labeled with $m=2$.
Next we notice that there exist irreducible multi-meson exchange diagrams
between the projectile and the target nucleus shown in Figures  \ref{figseven}d, \ref{figseven}e and \ref{figseven}f. 
Since our aim is to write
a multiple scattering theory similar to the non-relativistic theory, we need
to classify the diagrams in some way so that the kernel $V$ can be indexed
by the nucleon index.  For example, we can label the diagram
in Figure  \ref{figseven}d by $\sum_{ij} L_{ij}$ and Figure  \ref{figseven}e and \ref{figseven}f by
$\sum_{i,j\ne i} M_{ij}^{(1)}$ and $\sum_{i,j\ne i} M_{ij}^{(2)}$
etc. Now it is obvious that every diagram can be written
in the form $\sum\limits_i F_i$.
From experience with the non-relativistic
theory, we know that at a later point, we would like do the resummation of
the Born series in terms of a free $t$-matrix and in the relativistic
case, it might be a $t$-matrix calculated from some One Boson Exchange (OBE) 
model. Thus we can separate  $K_{0i}$  from the rest
of the terms in the kernel, as in
\begin{eqnarray}
V= \sum\limits_{i=0}^{A} V_{0i} =  \sum\limits_{i=0}^{A} ( K_{0i} + \Lambda_{0i} )  \label{relsum}
\end {eqnarray}
with
\begin{eqnarray}
\Lambda_{0i} \equiv \sum\limits_{i} \sum\limits_{m}  K_{ii}^{(m)} 
+\sum\limits_{i} \sum\limits_{m}  K_{iii}^{(m)} 
+ \sum\limits_{i,j} L_{ij} 
+ \sum\limits_{i,j\ne i} \sum\limits_{m} M_{ij}^{(m)}  + \cdots
\end {eqnarray}
{\bf } 
We  have separated $K_{0i}$ from the rest of the terms, but we could have chosen to either keep all terms or separate a particular subset of terms of interest. We will continue to study the separation of the OBE term  $K_{0i}$ in order to illustrate the technique.

Note that in $V$ we have separated $K_{0i}$ from the
other terms which we call $\Lambda_{0i}$. 
The $K_{0i}$ term is the OBE
term and $K_{ii},~K_{iii}$ etc. are two-meson, three meson exchange terms
respectively.  Depending 
on the phenomenological model, these contributions are sometimes modeled as
$\sigma$ exchange and other heavy meson exchanges.
The rest of the terms in $\Lambda_{0i}$ 
are diagrams where there can be more than one target nucleon
involved. The cross meson exchange diagram 
shown in Figure  \ref{figseven}d is where the projectile
exchanges two mesons with the target nucleus and in the intermediate state
the target is in some A-body excited state. In the non-relativistic theory
there is no such thing as a cross meson exchange, but in some crude 
way this type of 
diagram can be related to the nuclear correlation function in the 
non-relativistic theory.

\subsection{Relativistic optical potential}

We now define the projector 
to the target ground state $P$ and to the 
excited states $Q$. Assume that the A-body target bound state problem
can be solved in some way by employing methods such as the QHD \cite{dirk}
model. The labeling scheme is exactly
like the non-relativistic case. Therefore 
we can write the Bethe-Salpeter equation
as a coupled equation and define   the optical potential 
U as in the non-relativistic case,
\begin{eqnarray}
T&=&U+UPG_0PT  \label{bb1} \\
U&=&V+VQG_0QU  \label{bb2}
\end{eqnarray}
Now we are in a position to make a multiple scattering series for the
optical potential $U$. We first write
\begin{equation}
U=\sum\limits_{i=1}^A U_{0i}=\sum\limits_{i=1}^A  ( K_{0i}  +\Lambda_{0i} )
+\sum\limits_{i=1}^A ( K_{0i}  +\Lambda_{0i}  )  QG_0Q 
\sum\limits_{i=j}^A U_{0j} 
\end{equation}
Here we see great flexibility in formulating a multiple scattering theory.
The main aim in formulating a multiple scattering theory 
for the optical potential is to rewrite the series 
written in terms of fundamental interactions
into a series in terms of some scattering amplitudes. 
We have the flexibility in the sense that 
when we rewrite the series in terms of $t$-matrices, we can
 choose what we want for the $t$-matrix in the 
multiple scattering
series of the optical potential.
We have mentioned above that the $\Lambda_{0i}$ part contains diagrams
with two or more meson exchange between the 
projectile and the target.   
At this point we
can choose to include or not to include $\Lambda_{0i}$ or some part of
$\Lambda_{0i}$ in the kernel of
the $t$-matrix in the multiple scattering series of the optical potential. Since we want to formulate an RMST optical potential, whose first order single scattering term is given by the one boson exchange 
 free $t$-matrix, we will neglect the $\Lambda_{0i}$ terms. 
If we do not include the $\Lambda_{0i}$ terms in the $t$-matrix,
then following (\ref{tau}),  we can define 
\begin{equation}
{\hat \tau}_{0i} \equiv {K}_{0i} + {K}_{0i}{QG_0Q} {\hat \tau}_{0i}  \label{taudefine}
\end{equation}
and we  get a multiple scattering series for the optical potential as
  \begin{eqnarray}
U&=&\sum\limits_{i=1}^{A} {\hat \tau}_{0i} + \sum\limits_{i=1}^{A} 
{\hat \tau}_{0i}Q
G_0 Q\sum\limits_{j \ne i}^{A} U_{0j}
 +\sum\limits_{i=1}^{A} {f}_{0i} + \sum\limits_{i=1}^{A} {f}_{0i}Q
G_0Q\sum\limits_{j \ne i}^{A} U_{0j}  \label{reloptical} \\
&=&\sum\limits_{i=1}^{A} ({\hat \tau}_{0i} +{f}_{0i}) + \sum\limits_{i=1}^{A} 
({\hat \tau}_{0i} +{f}_{0i} )Q
G_0 Q\sum\limits_{j \ne i}^{A} U_{0j}
\end{eqnarray}
where $f_{0i}$ is defined as
\begin{eqnarray}
f_{0i}&\equiv &\Lambda_{0i} + K_{0i} QG_0Q f_{0i} \nonumber\\
&=&\Lambda_{0i}+K_{0i} QG_0Q \Lambda_{0i}+K_{0i} QG_0Q  K_{0i}QG_0Q\Lambda_{0i}
+\cdots \nonumber\\
&=&\Lambda_{0i}+{\hat \tau}_{0i} QG_0Q \Lambda_{0i}  \label{last}
\end{eqnarray}
The series given by equation   (\ref{reloptical})
is the relativistic multiple scattering series for the optical potential
in the Bethe-Salpeter formalism.  Compare  to the non-relativistic Watson optical potential in equation  (\ref{watsonoptical}).
The first term in    (\ref{reloptical})  is the single scattering 
term. The second term will produce, after iteration, the 
double scattering term etc. 
We have found that
there are diagrams in which the projectile is interacting with two or
more target nucleons via meson exchange. These terms are represented by the
terms with $f_{0i}$ in the second line of equation   (\ref{reloptical}).
 It is possible to include the $\Lambda_{0i}$
terms in the kernel
of the pseudo  two-body operator $\hat \tau$, but doing so will not give us any
advantages in approximating $\hat \tau$ by some suitable free two-body
Bethe-Salpeter amplitude at a later stage. We have to remember that the
main aim in formulating a multiple scattering series is to replace the
infinite series written in terms of fundamental interactions (such as OBE)
by a series in some two-body amplitude (such as free Bethe-Salpeter $t$-matrix)
which itself contains the fundamental interaction
to infinite order.

{\bf } The multiple scattering
series given by equation   (\ref{reloptical})  is formulated in four dimensions and 
{\em 
we have not yet made any approximation nor dimensional
reduction of any of the equations involved.} We have separated off the OBE term in order to illustrate how one might go about isolating particular terms of interest. However this separation does not involve any approximation because equations (\ref{taudefine}) - (\ref{last}) remain equivalent to the Bethe-Salpeter equation (\ref{bse}) together will all terms contained in (\ref{relsum}).
One could have separated off other terms in a similar manner. Or one might not separate off anything and keep the entire series, in which case none of the $f_{0i}$  terms would be present, and the ${\hat \tau}_{0i}$ term in (\ref{taudefine}) would 
instead read 
\begin{eqnarray}
{\hat \tau}_{0i} \equiv {V}_{0i} + {V}_{0i}{QG_0Q} {\hat \tau}_{0i}
\end{eqnarray}
just as in the non-relativistic case (\ref{tau}). However, again we continue to isolate the OBE terms in order to illustrate the technique. It is of interest to know the size of contribution of the crossed box diagram to the scattering amplitude in the Bethe-Salpeter equation. Although no one has done this within the context of the Bethe-Salpeter equation, Fleicher and Tjon \cite{fleicher} have analysed the relative sizes of the box diagram and the crossed box  diagram for on-shell  k-matrix-elements at 100 MeV. They found that the on-shell matrix elements for the crossed box are about 4 to 20 times weaker than their direct box  counterparts. They also noted that there exist some partial cancellations between the box and the crossed box diagrams.

\subsection{Relativistic impulse approximation}
\begin{figure}[!h]
\hspace*{3cm}  \includegraphics[width=12cm]{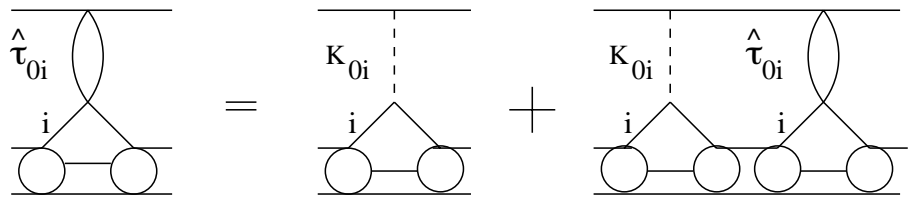}
\caption{The equation for the pseudo two-body operator
${\hat \tau}_{0i}$ which is the relativistic analogue of the
Watson $\tau_{0i}$ operator in the non-relativistic theory. }
\label{figeight}
\end{figure}

Just as in the non-relativistic case, we now have a multiple scattering
series for the optical potential. The series is written in terms of
a pseudo    two-body amplitude ${\hat \tau}_{0i}$ which has the effects of
many-body interaction in the kernel and  propagator. Because
solving ${\hat\tau}_{0i}$ involves all possible excited states of the
target, it is probably as hard as solving the original problem and for
any practical calculations we need to approximate this by the free two-body
amplitude. Before we make any approximation, we first examine the content
of this single scattering approximation to the optical potential. The single 
scattering optical potential is obtained by folding the
$\hat \tau$ amplitude with the target ground state, i.e.
$\langle \phi_0| \sum\limits_i {\hat \tau}_{0i} |\phi_0 \rangle$ and  the
equation for ${\hat \tau}_{0i}$  
is shown diagrammatically in Figure  \ref{figeight}. 
As in the non-relativistic case we do not want to calculate ${\hat \tau}_{0i}$
but  want to replace it in the multiple scattering series with a 
free two-body operator. The free two-body $t$ matrix is defined the same way as (\ref{free}), namely
 \begin{eqnarray}
\hat t^{\rm \, free}_{0i} \equiv \hat K_{0i} +\hat K_{0i} \hat g \hat t^{\rm \, free}_{0i}  
\end{eqnarray}
where $g$ is the free two body propagator, and the relation between $\hat t^{\rm \, free}$ and 
 $ \hat \tau$ is therefore
 \begin{eqnarray}
 \hat \tau_{0i} =  \hat t^{\rm \, free}_{0i} + \hat t^{\rm \, free}_{0i}(QG_0Q-\hat g) \hat \tau_{0i}  
\end{eqnarray}
analogous to  (\ref{relation}).

 {\bf } Note that we are introducing an approximation here because we are assuming that
  $\hat t^{\rm \, free}_{0i}$  involves only the OBE term $K_{0i}$ shown in
 Figure \ref{figseven}a. One might argue that this should also include the cross box term in
 Figure \ref{figseven}d, in which case one would repeat the above calculations, but separate off both the
  box (OBE) and cross box. All the equations above would then have $K_{0i}$ being defined as box (OBE) plus cross box, and the cross box term would be removed from $\Lambda_{0i}$. Nevertheless, for the sake of clarity,  we continue with separating only the OBE term.

\begin{figure}[!h]
\begin{center}
\includegraphics[width=7cm]{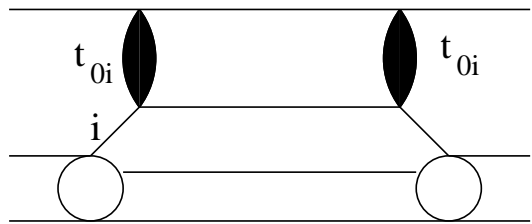}
 \end{center}
\caption{For medium energies one nucleon knockout terms
such as this dominate and the difference between
$\tau_{0i}QG_0Q \tau_{0i}$ and $t_{0i}QG_0Q t_{0i}$ is small.}
\label{fignine}
\end{figure}

Now we compare 
$\langle \phi_0| {\hat \tau} |\phi_0 \rangle$ and 
$\langle \phi_0| {\hat t^{\rm \, free}} |\phi_0 \rangle$.  Of course the 
difference between ${\hat \tau}_{0i}$ and ${\hat t^{\rm \, free}}_{0i}$ is the 
nuclear medium modification of the
interaction. But for intermediate and high energies where the impulse
approximation is good, the difference
is not significant. One contribution arising from medium modification is 
the shift in the energy of the terms in the kernel due to the motion of the
$A-1$ cluster. The second difference is in the iterated intermediate states 
where $\langle \phi_0| {\hat \tau} |\phi_0 \rangle$ includes excited
target intermediate states because of the propagator $QG_0Q$ in $\hat \tau$.
In order to see what is involved in approximating $\langle \phi_0| 
{\hat \tau} |\phi_0 \rangle$ by
$\langle \phi_0| {\hat t^{\rm \, free}} |\phi_0 \rangle$ we rewrite the optical potential
in terms of $\hat t^{\rm \, free}$,
\begin{eqnarray}
U&=&\sum\limits_{i=1}^{A} {\hat t^{\rm \, free}}_{0i} + \sum\limits_{i=1}^{A}
{\hat t^{\rm \, free}}_{0i}(QG_0Q-{\hat g}){\hat t^{\rm \, free}}_{0i} +\cdots \nonumber\\
&& +\sum\limits_{i=1}^{A} {\hat t^{\rm \, free}}_{0i}Q
G_0 Q\sum\limits_{j \ne i}^{A} {\hat t^{\rm \, free}}_{0j}+ \cdots \nonumber\\
&& +\sum\limits_{i=1}^{A} {f}_{0i} + \sum\limits_{i=1}^{A} {f}_{0i}Q
G_0Q\sum\limits_{j \ne i}^{A} {\hat t^{\rm \, free}}_{0j}+ \cdots  \label{relcorrection}
\end{eqnarray}
 Compare  to the non-relativistic expression   (\ref{correction}).
In equation (\ref{relcorrection}) the first term in the series $\sum\limits_{i=1}^{A} {\hat t^{\rm \, free}}_{0i}$ when
sandwiched between the target ground states will give the first-order single
scattering optical potential in the impulse approximation. 
The second term in the series is the propagator
correction term. In the non-relativistic theories, the 
name Impulse Approximation comes from the fact that
in medium and higher energies $\langle \phi_0|t_{0i}QG_0Qt_{0i}|\phi_0\rangle$
can be approximated well by $\langle \phi_0|t_{0i}gt_{0i}|\phi_0\rangle$ where
$g$ is the free two-body propagator. Obviously this will be a good 
approximation if  $\langle \phi_0|t_{0i}QG_0Qt_{0i}|\phi_0\rangle$  is dominated
by single nucleon knockout terms shown in Figure  \ref{fignine}. 

The second 
line in the above series (\ref{relcorrection})
are the double, triple, etc. scattering terms. Non-relativistically the first 
term plus the double scattering 
term constitute the second-order optical potential
in the impulse approximation. 

Diagrams 7e , 7f and other similar diagrams can be understood as three-body and multi-nucleon force terms in the nonrelativistic theory. Although it is possible to include them formerly in our two-body t-matrix, in order to see the OBE contribution and these multi-nucleon force terms separately, we lump all these non-OBE contributions in the $f_{0i}$ terms  in
equation   (\ref{reloptical}) and equation   (\ref{relcorrection}).
We will leave the labor of estimating the sizes and
 effects of these terms to future work. In any case, in order to
obtain an RMST whose leading term is given by an OBE $t$-matrix, we 
do not include them in the kernel of ${\hat \tau}_{0i}$.

\subsection{3-dimensional reduction}

The Bethe-Salpeter equation (\ref{bse})  can be reduced from 4 to 3 dimensions by writing it as a set of coupled equations
\begin{eqnarray}
T &=& K+K\widetilde G_0 T  \label{quasi} \\
K &=& V + V(G_0-\widetilde G_0) K
\end{eqnarray}
where $\widetilde G_0$ is a 3-dimensional propagator, which may be written in the general form \cite{MNK}
\begin{eqnarray}
\widetilde G_0 = -2\pi i \int \frac{ds^\prime}{s-s^\prime +i \eta} \, f(s^\prime , s)\, \delta^+(A)\, \delta^+(B)
\end{eqnarray}
where $s$ is the square of the total 4-momentum and $ f(s^\prime , s)$ is a function with the requirement that 
$f(s , s)=1$. $A$ and $B$ are arguments of the delta function which depend on 4-momentum \cite{MNK}. These $\delta$ functions are such that they fix a prescription for first component of 4-momentum, $k_0$, and thereby kill a $\int dk_0$ integral reducing the problem from 4 to 3-dimensions. This procedure is called a 3-dimensional reduction of the Bethe-Salpeter equation, resulting in the 3-dimensional   equation (\ref{quasi}). 
  There are infinitely many three-dimensional
reductions possible \cite{jackson}. The reduction is done by
using some delta functions and the equations obtained by this method
are commonly known as quasi-potential equations. Besides the quasi-potential
equations, there exist other covariant three-dimensional equations designed
to obey certain principles. For example, Phillips and Wallace 
have developed an equation which satisfies gauge invariance to any desired
order in the kernel  \cite{phillips}. Pascalutsa and Tjon have designed an equation
satisfying charge conjugation \cite{onebody}. More details can be found in reference \cite{MNK}.

So far the formulation of our RMST  is
entirely in four dimensions and no  dimensional reduction has been made.
In the four-dimensional formalism, the propagator for the elastic scattering
 equation  (\ref{bb1}), is $PG_0P$ where $G_0$ is the Bethe-Salpeter propagator for
the nucleon-nucleus system and $PG_0P$ tells us that the target
is propagating in its ground state. Apparently 
the nucleon-nucleus    scattering calculation
has never been done in full four dimensions. In actual calculations, for
proton-nucleus scattering, a fixed energy 
Dirac equation is used with scalar and vector potentials calculated from the
$t\rho$ approximation of the optical potential. 
Thus  one has made the assumption that the interaction is instantaneous. 
This means the target is infinitely heavy and the projectile moves in the
instantaneous potential of the target nucleus because a fixed energy Dirac 
equation is a three-dimensional one-body equation. 
That means in using the Dirac equation, one has made two 
approximations. First, the Bethe-Salpeter propagator 
of the nucleon-nucleus   system is replaced by
some three-dimensional two-body propagator. Second, a proper one-body
limit of the chosen three-dimensional two-body propagator is the 
Dirac propagator. To see what is involved, rewrite   (\ref{bb1}) and (\ref{bb2}) as the coupled
integral equations,
\begin{eqnarray}
T&=&{\widetilde U} +{\widetilde U}P{\widetilde G}_0PT  \label{mmm23}\\
{\widetilde U}&=&U+U(PG_0P-P{\widetilde G}_0P){\widetilde U}
\end{eqnarray}
 Obviously the difficulty level in solving for $\widetilde U$ is the 
same as solving the original 4-dimensional problem. In order to obtain a 3-dimensional
elastic scattering equation, we choose a 3-dimensional propagator
$P{\widetilde G}_0P$. All that is required to maintain
unitarity is that $P{\widetilde G}_0P$ has the same elastic cut as
$PG_0P$.   Of course in picking ${\widetilde G}_0$ we must specify how
the nucleon-nucleus  
relative energy variable is going to be handled so that equation  (\ref{mmm23}) will be
a three-dimensional equation. It should be clear that the relative energy 
prescription is entirely contained in ${\widetilde G}_0$ and 
$P{\widetilde G}_0P$ just tells us that the target is 
propagating in its ground state. Once the 3-dimensional propagator
${\widetilde G}_0$ is chosen, we have to use the same prescription
for fixing the relative energy in evaluating $U$.  In the   
nucleon-nucleus case, $U=\sum\limits_i {\hat t^{\rm \, free}}_{0i}$ contains $\hat t^{\rm \, free}$
whose propagator $\hat g$ is the Bethe-Salpeter propagator
of the projectile and a target nucleon.  An important conclusion of the present paper, is that 
 to  be consistent one
must use the same prescription in fixing the relative energy in $G_0$ 
and $\hat g$. For example, in nucleon-nucleus scattering, if we are going
to use a nucleon-nucleon $t$-matrix using the Blackenbeclar-Sugar 
propagator, the elastic scattering equation should also be the Blackenbeclar-Sugar
equation.

The final elastic scattering equation need not be a Dirac equation.
Making it a Dirac equation involves the assumption that the target nucleus
is infinitely heavy and that the proper one-body limit of the equation
corresponding to the three-dimensional equation with the propagator
$P{\widetilde G}_0P$ is the Dirac equation. In reality  no nucleus is 
infinitely heavy although it can be a good approximation 
for many heavy nuclei.  We note also that the correct one-body
limit can be easily incorporated in quasi-potential (three-dimensional)
or other types of two-body equations \cite{onebody}.

In the case of meson projectiles, there are three different masses 
involved; the mass of the meson, the mass of the nucleon and the mass of the
nucleus. Because of the mass difference between the meson and the nucleon, it 
is not suitable to use 3-dimensional quasi-potential equations which put
both particles equally on mass-shell and it is also not entirely justifiable to put
the nucleon on-mass-shell since the nucleon is not infinitely massive. In 
our opinion, the most suitable 3-dimensional equation to use for  the meson-nucleon
amplitude is the Proportionally Off-Mass-Shell equation \cite{MNK}. 
The propagator of this equation can be easily modified for boson-fermion or
fermion-fermion cases so it can be used for both the nucleon-nucleon  
and the nucleon-nucleus   propagators. 
The major advantage of this equation over other quasi-potential equations
 is that it adjusts
the off-shellness of the particles according to their masses. When
one of the particles is infinitely massive, it reduces to a one-body equation
and if the masses are equal, it treats the particles symmetrically and 
it reduces to an equation known as the 
Todorov equation \cite{todorov}. Obviously, this propagator can be used for both
mesonic and nucleonic projectiles and also for the projectile-taget 
propagation. It also gives us the added advantage that it automatically adjusts
itself to the masses involved because of the physically meaningful prescription
for fixing the relative energy. It would be interesting to see the use of this
proportionally off-mass-shell equation in nucleon-nucleus   
scattering in the future.

\section{Conclusions}

 We have formulated a relativisitic multiple scattering series for the 
optical potential in the the case of nucleon-nucleus scattering. 
As in reference  \cite{maung-gross}  we started from the fact that the nucleon-nucleus scattering
amplitude is given by an infinite series of meson exchange diagrams between
the projectile and the target. This infinite series can be written as an integral
equation (Bethe-Salpeter equation) if we include all projectile-target
irreducible diagrams in the kernel. In contrast to reference  \cite{maung-gross}  we do not consider the
cancellation of the box and the crossed box diagrams, but derived a multiple scattering
series without making any dimensional reduction. In the full 4-dimensional formalism,
neither the projectile nor the target is put on-mass-shell and we do not have the
problem of spurious singularities arising from putting an excited target on-mass-shell.
As expected, the RMST for the optical potential is very 
similar to the non-relativistic
counterpart. The only difference is the appearence of some extra terms arising from
diagrams with the projectile interacting with two or more nucleons via meson exchange.
We show that just like the non-relativistic  case, the single scattering first-order
impulse approximation optical potential operator is given by the
free two-body Bethe-Salpeter $t$-matrix summed over the target nucleon index. 

In this paper we discussed how to formulate a relativistic multiple
scattering theory for the optical potential in projectile-nucleus
scattering. We did not discuss about target recoil or the center of mass
motion of the A-body target. In practical calculations these things have
to be taken into account. One way to incorporate the A-body center of mass
motion is to use the Moller frame transformation factor \cite{moller}.
Mutiplying the nucleon-nucleon t-matrix (calculated in the nucleon-nucleon
center of mass frame) by this factor will produce the t-matrix to be used
appropriate for the nucleon-nucleus center of mass frame. In the optimal
factorization of the optical potential, recoil of the struck nucleon can be
taken into account by including a $- ({\bf p} + {\bf p^\prime})/2A$ term in the  struck
nucleon momenta where ${\bf p}$ and ${\bf p^\prime}$ are the initial and final
momentum in the nucleon-nucleus center of mass  frame and A is the mass number of the
target nucleus \cite{recoil}.  An in depth analysis of the effects of including
boost, recoil, Moller factor and Wigner rotation in proton-nucleus
scattering can be found in a study by Tjon and Wallace \cite{recoil}.
 
 We have discussed that there are many possible ways to organize the relativistic multiple scattering theory. Indeed, unlike the 
non-relativistic case, the relativistic case already has a kernel that includes multiple scattering at the level of meson exchange. One could in principle obtain a multiple scattering series which has the exact same form as the non-relativistic case (Eq. 45) by including these $f_0i$ diagrams in the definition of ${\hat \tau}_{0i}$. This shows that one can obtain a relativistic multiple scattering series for the optical potential in the mold of the non-relativistic theory. As far as we are aware, all relativistic nucleon-nucleus scattering calculations that use a two-body t-matrix calculated from a two-body relativistic equation have used OBE models. Therefore we keep the OBE contribution and contributions from the many-body force diagrams separate so that we can see what is left out in these calculations. In this paper we try to stay close to the non-relativisitc Watson formalism. In the literature on the non-relativistic multiple scattering theory there are other ways to organize the multiple scattering series \cite{kmt, siciliano, ernst}. Developing such organizations are beyond the scope of this work.

{\bf } Throughout the paper, we have illustrated our technique by separating off the OBE term shown in 
Figure \ref{figseven}a. We have mentioned several times that this particular separation is not necessary, and we have discussed how to choose alternatives.  The use of the OBE term alone might be  a popular choice and our discussion  shows what approximations are involved in making such a choice and what terms are left out. 

Next we rewrote the elastic scattering equation into coupled integral equations by
introducing an auxilary interaction $\widetilde U$ and a propagator
$P{\widetilde G}_0 P$. This propagator contains a prescription for fixing the
relative energy variable and must also have the same elastic cut as $PG_0P$
so that it will obey unitarity. The final elastic scattering equation is
$T={\widetilde U} +{\widetilde U}P{\widetilde G}_0PT$ which is a
three-dimensional covariant equation. The 3-dimensional optical potential
$\widetilde U$ is obtained from $U$ by using the same relative energy prescription
as in $P{\widetilde G}_0 P$. This requires that the free two-body
$t$-matrix in the optical potential should be calculated with the same
relative energy prescription. To give a concrete example, if $P{\widetilde G}_0 P$
corresponds to Blackenbeclar-Sugar propagator, then the first order impulse
approximation optical optential is $\sum\limits_i {\hat t^{\rm \, free}}_i$ where ${\hat t^{\rm \, free}}_i$
must be calculated from the Blackenbeclar-Sugar equation. An important conclusion of 
this paper is that  the propagators of the elastic scattering equation
and the free two-body $t$-matrix must be consistent. 
Next, we have looked at what 
approximation is
involved in using a fixed energy Dirac equation. Obviously, from the discussion above
the final projectile-target elastic scattering equation does not have to be a Dirac
equation. Since the fixed energy Dirac equation is a one-body equation, the use of it
implies that the target is infinitely heavy. The more subtle point involved here is
that, in doing so, we are also assuming that the correct one-body limit of
the elastic scattering two-body equation with propagator $P{\widetilde G}_0 P$
is a Dirac equation.  The effects of other propagators other than Dirac should be 
tested in future calculations, although we believe that 
for heavy target nuclei such as $^{40}$Ca or
$^{208}$Pb, Dirac RIA would be an excellent approximation, 
Finally, we argued that it is physically more meaningful and aesthetically
pleasing to use
the Proportionally Off-Mass-Shell propagator \cite{MNK} for projectile-nucleon
and nucleon-nucleus propagators regardless of whether the projectile is
a meson or a nucleon.

\noindent Acknowledgements: 
 KMM and TC would like to acknowledge the support of 
COSM, NSF Coorperative Agreement 
PHY-0114343 and Hampton University where part of this work was done. 
JWN  was supported by  NASA grant NNL05AA05G.

\end{document}